\documentclass[aps,physrev,reprint,superscriptaddress,longbibliography]{revtex4-2}

\usepackage{preamble}

\begin{document}

\title{\textbf{Near-field contributions to cavity-mediated radiative heat transfer}}

\author{Antonio Mastropasqua}
\affiliation{Department of Physics, Friedrich-Alexander-Universität Erlangen-Nürnberg, 91058 Erlangen, Germany}

\author{Martin Eckstein}
\affiliation{Department of Physics, Universität Hamburg, 22607 Hamburg, Germany}
\affiliation{The Hamburg Centre for Ultrafast Imaging, Hamburg, Germany}

\author{Daniele Fausti}
\affiliation{Department of Physics, Friedrich-Alexander-Universität Erlangen-Nürnberg, 91058 Erlangen, Germany}
\affiliation{Department of Physics, Università degli Studi di Trieste, 34127 Trieste, Italy}

\author{Francesca Fassioli}
\affiliation{Department of Physics, Friedrich-Alexander-Universität Erlangen-Nürnberg, 91058 Erlangen, Germany}

\begin{abstract}
    Embedding  materials in optical cavities has emerged as a powerful strategy for controlling their physical properties. In this context, increasing attention has been devoted to the \emph{thermal Purcell effect}, which describes how the radiative heat transfer between a material and its photonic environment can be modified by the electromagnetic boundary conditions imposed by the cavity. Here, we investigate \emph{near-field effects} in cavity-mediated radiative heat transfer in planar cavities composed of realistic, dissipative gold mirrors, extending previous work that considered ideal mirrors. We identify the conditions under which  near-field effects lead to a substantial enhancement of heat transfer between the material and a photonic thermal bath relative to free space, owing to the strong modification of the photonic local  density of states by evanescent modes within the cavity. In particular, we find that radiative heat flow can be enhanced by several orders of magnitude in cavities ranging from nanometer to micrometer scales, depending on the absorber frequency and dipole orientation. These findings suggest that near-field effects may be relevant to polariton chemistry experiments and other systems operating within this cavity length-scale regime. Our results add to the idea that cavities can modify material properties not only by engineering  vacuum fluctuations in the strong light-matter coupling regime, but also by controlling radiative dissipation and heat transfer.
\end{abstract}

\maketitle

\section{Introduction}
\label{sec:introduction}
\begin{figure}
    \centering
    \begin{subfigure}{\columnwidth}
        \caption{}
        \begin{tikzpicture}

    \def\ox{-3}
    \def\oy{4.5}
    % origin
    \coordinate (O) at (\ox,\oy);
    
    % axes
    \draw[->, thick] (O) -- (\ox + 0.5,\oy) node[anchor=west] {$\hat{z}$};
    \draw[->, thick] (O) -- (\ox,\oy-0.5) node[anchor=north east] {$\hat{y}$};
    
    % x axis out of plane (dot)
    \draw (O) circle (2pt) node[anchor=south east] {$\hat{x}$};
    \filldraw (O) circle (0.8pt);

    \def\h{3.5}

    \draw[<->, thick] (-1.5, \h + 0.2) -- (1.5, \h + 0.2) node[midway, above] {L};

    \shade[shading=axis, right color=gold!80!black, middle color=gold!30!white, left color=white] (-3,0) rectangle (-1.5,\h);
    \node at (-2.25, \h/2) {$T_{\mathrm{gold}}$};

    \shade[shading=axis, left color=gold!80!black, middle color=gold!30!white, right color=white] (1.5,0) rectangle (3,\h);
    \node at (2.25, \h/2) {$T_{\mathrm{gold}}$};

    % \shade[shading=axis, top color=darkred, middle color=lightred, bottom color=red] (-0.5,0) rectangle (0.5,\h);
    % \node at (0, 0.3) {$T_m$};
    \shade[shading=axis, top color=armygreen, middle color=camouflagegreen, bottom color=pistachio] (-0.5,0) rectangle (0.5,\h);
    \node at (0, 0.3) {$T_m$};

    \def\xstartR{0.5}
    \def\xendR{1.5}
    \def\xstartL{-1.5}
    \def\xendL{-0.5}
    \pgfmathsetmacro{\L}{\xendR-\xstartR}
    \def\A{0.3}
    \pgfmathsetmacro{\yshiftU}{\h - 0.9}
    \def\yshiftD{0.5}
    \def\lam{4}

    \draw[->, thick, color = blue] (\xstartR, \yshiftU + 0.1) -- (\xendR, \yshiftU + 0.1);
    \draw[->, thick, color = blue] (\xendR, \yshiftU - 0.1) -- (\xstartR, \yshiftU - 0.1);

    \draw[->, thick, color = blue] (\xendL, \yshiftU + 0.1) -- (\xstartL, \yshiftU + 0.1);
    \draw[->, thick, color = blue] (\xstartL, \yshiftU - 0.1) -- (\xendL, \yshiftU - 0.1);

    \draw[thick, color = blue, domain=\xstartR:\xendR, samples=200] plot (\x, {\yshiftU + \A*sin(4*pi*(\x-\xstartR)/\L r)});
    \draw[thick, color = blue, domain=\xstartL:\xendL, samples=200] plot (\x, {\yshiftU + \A*sin(4*pi*(\x-\xstartL)/\L r + pi r)});

    \draw[thick, color=red, domain=\xstartR:\xendR, samples=200] plot (\x, {\yshiftD + exp(-\lam*(\xendR-\x))});
    \draw[-, thick, color=red] (\xstartR, \yshiftD) -- (\xendR, \yshiftD);
    \draw[thick, color=red, domain=\xstartL:\xendL, samples=200] plot (\x, {\yshiftD + exp(-\lam*(\x-\xstartL))});
    \draw[-, thick, color=red] (\xendL, \yshiftD) -- (\xstartL, \yshiftD);

    \foreach \x in {0.6,0.8,1.0,1.2,1.4} {\draw[red, thick] (\x, \yshiftD)-- (\x, {\yshiftD + exp(-\lam*(\xendR-\x))});}
    \foreach \x in {-1.4,-1.2,-1.0,-0.8,-0.6} {\draw[red, thick] (\x, \yshiftD)-- (\x, {\yshiftD + exp(-\lam*(\x-\xstartL))});}

    \node[blue, font=\bfseries] at (1,\yshiftU+0.5) {prop.};
    \node[blue, font=\bfseries] at (-1,\yshiftU+0.5) {prop.};
    \node[red, font=\bfseries] at (1,\yshiftD-0.2) {evan.};
    \node[red, font=\bfseries] at (-1,\yshiftD-0.2) {evan.};

\end{tikzpicture}
        \label{fig:sketch}
    \end{subfigure}
    \hfill
    \begin{subfigure}[t]{\columnwidth}
        \caption{}
        \includegraphics[width=\columnwidth]{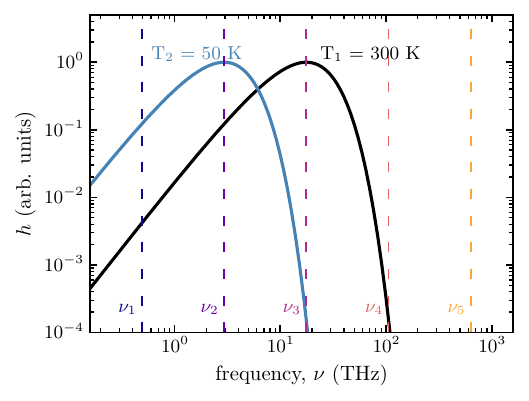}
        \label{fig:black_body}
    \end{subfigure}
    \caption{(a) A material at temperature $T_m$ is embedded in a cavity made of two semi-infinite gold mirrors at temperature $T_{\mathrm{gold}}$. (b) Spectral energy density $h(2 \pi \nu, T_l, T_m \to 0)$ as a function of the frequency $\nu$ for two blackbodies at temperatures $T_l = T_1 = 300$ K and $T_l = T_2 = 50$ K. The peak of the spectral energy density of a blackbody at temperature $T_l$ occurs at a frequency $\nu_{max}\approx 2.82 k_BT_l/(2\pi\hbar)$. The vertical lines show five relevant log-equally spaced frequencies, with the central one $\nu_3= \nu_{max}(T_l = T_1 = 300\, \mathrm{K}) \approx 17.6$ THz.}
\end{figure}

The engineering of \emph{strong light-matter interactions} using optical cavities to modify material functionalities has emerged as a central theme in condensed matter physics and chemistry~\cite{Garcia-Vidal, Schlawin, Ebbesen}. In the strong coupling regime, coherent energy exchange between light and matter dominates over dissipative processes, leading to the formation of hybrid light–matter states. This provides a route to modify the properties and functionality of molecules and materials through their photonic environment. \\
\indent Importantly, the macroscopic properties of matter can also be modified by light-matter interactions in the \textit{weak light-matter coupling} regime. In contrast to strong-coupling, the weak-coupling regime is characterized by the absence of coherent hybridization between light and matter. Instead, the primary effect of the cavity is to modify the electromagnetic mode structure, reshaping the local photonic environment in which matter resides and, thereby, the way in which energy is exchanged between matter and light. This modification is quantitatively captured by the partial local density of states (PLDOS), denoted by $\rho(\hat{\bm{e}}_a, \bm{r}, \omega)$, which describes the density of states of the electromagnetic field at a frequency $\omega$, position $\bm{r}$, and along the direction $\hat{\bm{e}}_a $~\cite{Barnes}. In free space, the PLDOS follows a $\propto \omega^2$ frequency scaling and is independent of position and direction, but in the presence of a cavity, boundary conditions imposed on the electromagnetic field lead to strong spatial, directional and spectral restructuring. Such modifications of the electromagnetic environment can enhance or suppress radiative processes, as first recognized by Purcell in the context of spontaneous emission~\cite{Purcell}. \\
\indent In the context of thermal physics, the cavity-modified photon environment has direct consequences for radiative heat transfer. Motivated by the experimental demonstration of a cavity-mediated thermal control of the metal-to-insulator transition in $1$T$-$TaS$_2$~\cite{Jarc}, we recently proposed~\cite{Fassioli} that cavities can modify the radiative heat exchange between a material and an external thermal photon bath, thereby controlling the material temperature when this is also in contact with a second thermal source (e.g. a cold finger). In particular, we showed that the modification of the PLDOS by a cavity composed of ideal mirrors can appreciably alter the radiative heat flow relative to free space when the cavity fundamental frequency is tuned to the spectral region where the free-space radiative heat flow is largest. At ambient temperature, this condition corresponds to the mid-IR regime (micrometer length-scale cavities). Similar conclusions were drawn by~\cite{Chiriaco} and~\cite{Pannir}. \\
\indent The aforementioned works considered either perfect mirrors~\cite{Fassioli, Chiriaco} or only the propagating part of the electric field in imperfect mirrors~\cite{Pannir}. Realistic (imperfect) mirrors not only confine propagating modes but also, importantly, support a continuum of evanescent electromagnetic modes which decay exponentially with the distance from the mirror surface (see~\cref{fig:sketch}). 
At short distances from the mirror, $d \ll \lambda$, these evanescent modes become increasingly important and can dominate electromagnetic interactions. This evanescent contribution to the electromagnetic field is well known to play a central role in near-field radiative heat transfer.
When two objects are brought to separations smaller than the thermal wavelength $\lambda_{\mathrm{th}}= \hbar c/(k_B T)$, evanescent electromagnetic modes can tunnel between the two, providing an additional channel for heat transfer and giving rise to the so-called near-field enhancement. In contrast, at separations larger than $\lambda_{\mathrm{th}}$, radiative heat exchange between two objects is governed by far-field effects and is bounded by Planck's blackbody radiation law. Near-field enhancement in radiative heat transfer was first theoretically predicted by Polder and Van Hove~\cite{Polder}, and has been subsequently investigated in detail across a broad range of material systems and geometries through both theoretical studies~\cite{Chapuis, Cuevas, Mulet, Volokitin, Pendry, Principi, Biehs, Ben-Abdallah} and experimental works~\cite{Kittel, Rousseau, Shen, Kim, Song, Kloppstech, St-Gelais}.
This evanescent enhancement is well established in radiative heat transfer between two closely spaced parallel plates or between nanoparticles and a metallic interface~\cite{Barnes, Carminati, Joulain, Chapuis, Biehs, Ben-Abdallah, Chapuis_nanoparticles, Cuevas}. However, its role in radiative heat exchange between a material embedded in a Fabry-Pérot cavity and its photonic environment has not yet been explored for realistic, dissipative  metallic boundaries.\\
\indent In this work, we study near-field effects on the radiative heat transfer between a material and its electromagnetic environment when the material is embedded in a planar cavity formed by two semi-infinite gold mirrors. We extend the framework of~\cite{Fassioli} to realistic dissipative metallic mirrors that support evanescent modes. We employ the electric dyadic Green's function formalism~\cite{Barnes, Novotny} to derive analytical expressions for the propagating and evanescent contributions to the PLDOS inside the cavity for both parallel and perpendicular (to the mirror plane) dipole orientations, valid for arbitrary cavity lengths and frequencies. We then use these expressions to evaluate radiative heat transfer to a Lorentz-oscillator absorber and identify the regimes of cavity length, dipole orientation, and oscillator frequency in which near-field evanescent modes lead to significant deviations from the free-space heat exchange between the material and the photon bath. The paper is organized as follows. In~\cref{sec:model} we introduce the model for heat transfer, where the heat transfer rate can be directly related to the PLDOS, and derive the key expressions for the PLDOS. In~\cref{sec:PLDOS} we analyze the propagating and evanescent contributions to the PLDOS as functions of frequency and cavity length. In~\cref{sec:RHT} we present results for the radiative heat transfer, and in~\cref{sec:conclusions} we discuss our findings and their implications.

\section{Model}
\label{sec:model}
\subsection{Radiative heat power}
Here we introduce the model for radiative heat transfer between light and a material in terms of the PLDOS, following the same procedure as in Ref.~\cite{Fassioli}. The radiative heat flow between an electromagnetic field and a material is given by
\begin{equation}
\label{Qdot}
    \dot{Q} = - \int_V d^3r\, \langle \dot{\hat{d}}(\bm{r}) \hat{E}(\bm{r}) \rangle\,,
\end{equation}
where $\hat{E}(\bm{r})$ is the electric field and $\hat{d}(\bm{r})$ is the electric polarization density of the material. In~\cref{Qdot} we have implicitly assumed that the material considered has no magnetic polarizability such that contributions to the radiative heat flow from the magnetic field have been neglected. Under weak light-matter coupling and assuming that both the radiation field and matter can be described by a thermal state at temperatures $T_l$ and $T_m$ respectively, the expectation value of the heat rate in~\cref{Qdot} can be expressed as~\cite{Fassioli}, 
\begin{align}
\label{Qdot2}
\langle \dot Q \rangle 
&= 
\sum_{a,b=x, y, z}\int \frac{d\omega}{\pi}
\int d^3r\, d^3r'
D_{a,b}''(\bm{r},\bm{r}';\omega)
\alpha''_{b,a}(\bm{r}',\bm{r};\omega) \nonumber\\
&\times\,\,\,\,
\hbar\omega
\left(
b(\omega,T_l)
-
b(\omega,T_m)
\right)\,.
\end{align}
$D_{a,b}''({\bm r},{\bm r}\,';\omega)$ is the imaginary part of the Fourier-transformed retarded propagator of the electric field $D_{a,b}(\bm{r},\bm{r}';t-t') = \frac{1}{i\hbar}\theta(t-t') \big\langle [\hat E_a(\bm{r},t),\hat E_{b}(\bm{r}',t')] \big\rangle$, $\alpha''_{b,a}(\bm{r}',\bm{r};\omega)$ is the imaginary part of the material's polarizability density, and $b(\omega, T)=1/(e^{\hbar\omega/kT}-1)$ is the Bose-Einstein distribution at temperature $T$. \\
\indent The radiative heat power in~\cref{Qdot2} depends, in general, on spatial correlations of the electric field $\langle \hat E_a(\bm{r},t) \hat E_{b}(\bm{r}',t') \rangle$ between two positions $\bm{r}$, $\bm{r}'$ in the sample. If we now make the further assumption that the polarizability is local, $\alpha(\bm{r'},\bm{r}, \omega)\sim \delta(\bm{r} - \bm{r'}) \alpha(\bm{r}, \omega)$, i.e. the polarization at position $\bm{r}$ depends only on the local electric field $\hat{\bm{E}}({\bm{r})}$, then the heat flow becomes an integral over the local heat flow density $\dot{q}(\bm{r})$ at position $\bm{r}$, $\dot{Q}= \int_V d^3r\ \dot{q}(\bm{r})$. This is a fundamental requirement in order to write the heat flow in terms of the PLDOS as we will do next. For simplicity, we choose the principal axis of the polarizability to be along the x, y, z coordinates, such that $D_{a,b}''(\bm{r},\bm{r};\omega) = \delta_{a,b} D_{a,a}''(\bm{r},\bm{r};\omega)$, and note that $D_{a,a}''(\bm{r},\bm{r};\omega) = - \pi\omega \rho(\hat{\bm{e}}_a, \bm{r}, \omega) /(2\epsilon_0)$, where $\rho(\hat{\bm{e}}_a, \bm{r}, \omega)$ is the PLDOS. We can finally write the heat flow per unit volume as 
\begin{equation}
\label{eq:heat_flow}
    \dot{q}(\bm{r}) = \sum\limits_{a = x,y,z} \int_0^{\infty} d\omega\, \frac{\rho(\hat{\bm{e}}_a, \bm{r}, \omega)}{\rho_{\mathrm{free}}(\omega)}\, \sigma_{aa}(\bm{r},\omega)\, h(\omega, T_l, T_m)\,,
\end{equation}
with
\begin{equation}
\label{h_w}
    h(\omega,T_l,T_m) = \frac{\hbar \omega \rho_{\mathrm{free}}(\omega)}{\epsilon_0} \left( b(\omega,T_l) - b(\omega,T_m) \right)\,.
\end{equation}
In this final expression, we have replaced the polarizability by the real part of the optical conductivity, $\sigma(\omega)=-\omega \alpha''(\omega)$, and $\rho_\mathrm{free}(\omega)= \omega^2/(3 \pi^2 c^3$) is the free-space PLDOS. Note that $h(\omega,T_l,T_m \to 0)$ corresponds to the spectral energy density of a blackbody ($\sigma=constant$) at temperature $T_l$. As can be seen from~\cref{eq:heat_flow}, the integrand is directly proportional to the PLDOS normalized to free-space $\rho(\hat{\bm{e}}_a, \bm{r}, \omega)/\rho_\mathrm{free}(\omega)$ weighted by the conductivity of the sample $\sigma(\omega)$ and the energy density $h(\omega,T_l,T_m)$. The latter sets the frequency range at which the heat transfer is maximized in free space, which corresponds to the IR regime at room temperature. As we will show in subsequent sections, the evanescent contribution to the PLDOS in imperfect cavities can lead to substantially greater enhancement of the heat transfer than is possible in idealized cavities. Whereas the latter enhancement requires the cavity fundamental frequency to be tuned to the spectral region that dominates heat transfer in free space, the evanescent contribution can produce a strong enhancement when the cavity length is instead much shorter than the corresponding wavelength range.

\subsection{PLDOS inside imperfect planar cavity}
We consider a planar cavity formed by two semi-infinite gold layers separated by a distance $L$ along the z-direction with a cavity fundamental frequency of $\nu_c = c/(2L)$ (\cref{fig:sketch}). 
In contrast to the approach adopted in~\cite{Fassioli}, where the PLDOS is expressed as a sum over cavity modes $\rho(\hat{\bm{e}}_a, \bm{r}, \omega) = \sum_m \left| \hat{\bm{e}}_a \cdot \bm{\mathcal{E}}_m(\bm{r}) \right|^2 \delta(\omega - \omega_m)$, with $\bm{\mathcal{E}}_m$ the electric field modes and $\omega_m$ the corresponding eigenfrequencies, such a decomposition is not applicable in the presence of dissipative materials, such as gold. The reason is that the corresponding modes no longer possess purely real eigenfrequencies, so that the Dirac delta function is not well defined. Instead, we employ the more general formulation based on the electric dyadic Green's function~\cite{Chew}
\begin{equation}
\label{eq:defrho}
    \rho(\hat{\bm{e}}_a, \bm{r}, \omega) = \frac{2\omega}{\pi c^2} \hat{\bm{e}}_a \cdot \text{Im}\left[\mathbb{G}(\bm{r}, \bm{r}, \omega) \right] \cdot \hat{\bm{e}}_a\,.
\end{equation}
The electric dyadic Green's tensor $\mathbb{G}(\bm{r}, \bm{r'}, \omega)$ is related to the time-harmonic electric field via $\bm{E}(\bm{r},\omega) = i \omega \mu \mu_0 \int_{V'} d^3r'\, \mathbb{G}(\bm{r}, \bm{r}', \omega) \cdot \bm{j}(\bm{r}')$. This function describes how the electric field at a point $\bm{r}$ responds to an electric dipole $\bm{j}$ at the source point $\bm{r'}$~\cite{Novotny, Tai, Dung}. The formulation of electromagnetic fluctuations in terms of the dyadic Green's function and the corresponding local density of states is well established in fluctuational electrodynamics~\cite{Barnes, Joulain_2003}. Here we apply this framework to the PLDOS relevant for heat exchange of a material embedded in a dissipative Fabry--Pérot cavity, explicitly separating propagating and evanescent contributions. \\
\indent In a planar cavity geometry, it is natural to separate the wavevector $k_0=\omega/c$ into its parallel $k_{\parallel}$ and perpendicular $k_z$ components. Propagating modes correspond to $k_{\parallel} \le k_0$, whereas evanescent modes satisfy $k_{\parallel} > k_0$. Using~\cref{eq:defrho} (see Supplemental Material and~\cite{Barnes} for the detailed derivation), we calculate the propagating and evanescent contributions to the PLDOS for a dipole located at the center of the cavity ($z = L/2$), and for dipole orientations parallel ($\hat{\bm{e}}_a = \hat{\bm{e}}_x$) and perpendicular ($\hat{\bm{e}}_a = \hat{\bm{e}}_z$) to the plane of the cavity mirrors. The propagating contributions are given by 
\begin{widetext}
    \begin{subequations}\label{eq:rhoprop}
        \begin{align}
            \frac{\rho^{\parallel}_{\mathrm{prop}}(\hat{\bm{e}}_x, z = L/2, \omega)}{\rho_{\mathrm{free}}(\omega)} &= 1 + \frac{3}{2 k_0} \int_{0}^{k_0} dk_{\parallel} \frac{k_{\parallel}}{k_z} \frac{\text{Re} \left[r_1 e^{i k_z L}\right] - \left| r_1 \right|^2  }{\left| 1 - r_1 e^{i k_z L} \right|^2} - \frac{3}{2 k_0} \int_{0}^{k_0} dk_{\parallel} \frac{k_{\parallel} k_z}{k_0^2}  \frac{\text{Re} \left[r_2 e^{i k_z L}\right] + \left| r_2 \right|^2  }{\left| 1 + r_2  e^{i k_z L} \right|^2}\,, \label{eq:rhoprop1} \\
            \frac{\rho^{\perp}_{\mathrm{prop}}(\hat{\bm{e}}_z, z = L/2, \omega)}{\rho_{\mathrm{free}}(\omega)} &= 1 + \frac{3}{k_0} \int_{0}^{k_0} dk_{\parallel} \frac{k_{\parallel}^3}{k_0^2 k_z} \frac{\text{Re} \left[r_2 e^{i k_z L}\right] - \left| r_2 \right|^2  }{\left| 1 - r_2 e^{i k_z L} \right|^2}\,, \label{eq:rhoprop2}
        \end{align}
    \end{subequations}
while the evanescent ones read
    \begin{subequations}\label{eq:rhoevan}
        \begin{align}
        \frac{\rho^{\parallel}_{\mathrm{evan}}(\hat{\bm{e}}_x, z = L/2, \omega)}{\rho_{\mathrm{free}}(\omega)} &= \frac{3}{2 k_0} \int_{k_0}^{+\infty} dk_{\parallel} \frac{k_{\parallel}}{\mathcal{K}_z} e^{-\mathcal{K}_z L} \frac{\text{Im} \left[r_1 \right]}{\left| 1 - r_1 e^{- \mathcal{K}_z L} \right|^2} + \frac{3}{2 k_0} \int_{k_0}^{+\infty} dk_{\parallel} \frac{k_{\parallel} \mathcal{K}_z}{k_0^2} e^{-\mathcal{K}_z L} \frac{\text{Im} \left[r_2\right]}{\left| 1 + r_2 e^{- \mathcal{K}_z L} \right|^2}\,, \label{eq:rhoevan1} \\
        \frac{\rho^{\perp}_{\mathrm{evan}}(\hat{\bm{e}}_z, z = L/2, \omega)}{\rho_{\mathrm{free}}(\omega)} &= \frac{3}{k_0} \int_{k_0}^{+\infty} dk_{\parallel} \frac{k_{\parallel}^3}{k_0^2 \mathcal{K}_z} e^{-\mathcal{K}_z L} \frac{\text{Im} \left[r_2 \right]}{\left| 1 - r_2  e^{- \mathcal{K}_z L} \right|^2}\,. \label{eq:rhoevan2}
    \end{align}
    \end{subequations}
\end{widetext}
The integral is performed over all possible in-plane wavevectors $k_{\parallel}$. The perpendicular component of the wavevector $k_z =\smash[t]{(k_0^2 - k_{\smash{\parallel}}^2)^{1/2}}$ becomes purely imaginary ($k_z = i \mathcal{K}_z$) when $k_{\smash{\parallel}} > k_0$, leading to the exponentially decaying evanescent fields $\sim \smash{e^{-\mathcal{K}_z L}}$. These contributions are therefore strongly suppressed at large cavity separations ($L \gg \lambda$), but become dominant at subwavelength distances ($L \ll \lambda$). The PLDOS expressions involve the Fresnel reflection coefficients for the transverse electric ($r_1$) and transverse magnetic ($r_2$) polarizations at the vacuum-gold interface, given by
\begin{subequations}
    \begin{align}
    r_1(\omega,k_{\parallel}) &= \frac{\sqrt{k_0^2 - k_{\parallel}^2} - \sqrt{\epsilon^{\mathrm{Au}}(\omega) k_0^2 - k_{\parallel}^2}}{\sqrt{k_0^2 - k_{\parallel}^2} + \sqrt{\epsilon^{\mathrm{Au}}(\omega) k_0^2 - k_{\parallel}^2}}\,, \\
    r_2(\omega,k_{\parallel}) &= \frac{\epsilon^{\mathrm{Au}}(\omega) \sqrt{k_0^2 - k_{\parallel}^2} - \sqrt{\epsilon^{\mathrm{Au}}(\omega) k_0^2 - k_{\parallel}^2}}{\epsilon^{\mathrm{Au}}(\omega) \sqrt{k_0^2 - k_{\parallel}^2} + \sqrt{\epsilon^{\mathrm{Au}}(\omega) k_0^2 - k_{\parallel}^2}}\,.
    \end{align}
\end{subequations}
The dielectric response of gold is modeled through a Drude dielectric function $\epsilon^{\mathrm{Au}}(\omega) = 1 - \omega_p^2 / (\omega^2 + i \gamma \omega)$, with parameters $\smash{\omega_p = 1.71 \times 10^{4}}$ THz and $\smash{\gamma = 4.05 \times 10^{1}}$ THz~\cite{Chapuis}. Ref.~\cite{Chapuis} also investigated nonlocal effects on radiative heat flux between gold surfaces and found that local and nonlocal models yield comparable results for gap sizes larger than a few nanometers. Since the cavity separations considered here are above this length scale, the local-response approximation adopted in the Drude model is justified.\\
\indent It is important to note that the evanescent PLDOS contains terms proportional to the imaginary part of the reflection coefficients, and therefore there are no evanescent contributions to the PLDOS for purely real $r_1$ and $r_2$ (perfect mirrors). Also, since these coefficients are functions of the frequency $\omega$ through the Drude dielectric response and of the in-plane momentum $k_{\parallel}$, the PLDOS depend separately on $\omega$ and $\omega_c = 2 \pi \nu_c = \pi c/L$, rather than solely on their ratio $\omega/\omega_c$ as in the case of ideal mirrors.

\section{PLDOS in gold cavity: propagating and evanescent contributions}
\label{sec:PLDOS}
In order to investigate near-field effects on radiative heat flow, we first characterize the PLDOS inside a planar gold cavity and discuss the regime in which evanescent modes become relevant.

\subsection{PLDOS vs frequency}

We characterize the PLDOS as a function of frequency for five representative cavity lengths, chosen such that the fundamental frequency of the cavity matches the five frequencies $\nu_i$ with $i=\{ 1, 2, 3, 4, 5 \}$ in~\cref{fig:black_body}, where $\nu_3$ corresponds to the maximum of the blackbody spectrum at room temperature (T$_1$ = 300 K). The corresponding cavity lengths $L_i=c/(2\nu_i)$ are: $L_1 = 306.12\, \mathrm{\mu}$m ($\nu_1 = 0.49$ THz), $L_2 = 51.02\, \mathrm{\mu}$m ($\nu_2 = 2.94$ THz), $L_3 = 8.50\, \mathrm{\mu}$m ($\nu_3 = 17.6$ THz), $L_4 = 1.42\, \mathrm{\mu}$m ($\nu_4 = 105.77$ THz) and $L_5 = 0.24\, \mathrm{\mu}$m ($\nu_5 = 634.60$ THz). \\
\begin{figure*}[t]
    \centering
    \includegraphics[width=1.0\textwidth]{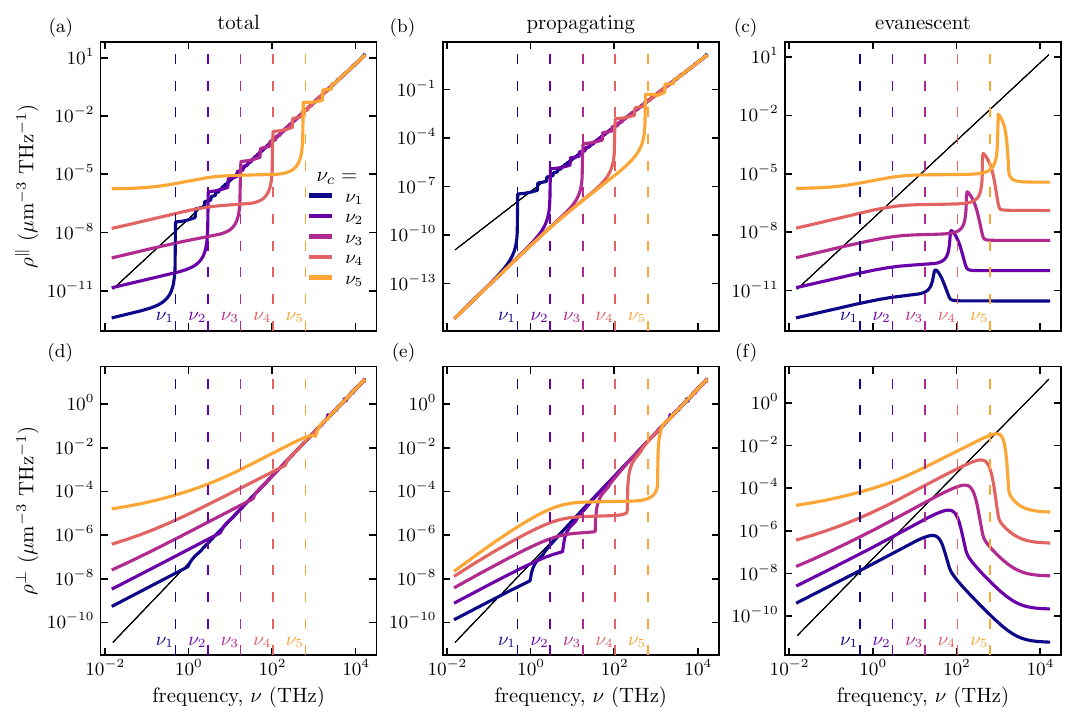}
    \caption{PLDOS vs frequency for different dipole orientations. (a), (d) Total, (b), (e) propagating and (c), (f) evanescent PLDOS for a parallel (upper row) and a perpendicular (lower row) dipole orientation, evaluated at the center of the cavity, as functions of frequency, where $\rho^{\parallel/\perp} = \rho(\hat{\bm{e}}_{x/z}, z = L/2, 2 \pi \nu)$. The five curves correspond to five different cavities with a cavity fundamental frequency $\nu_c$ matching the five frequencies $\nu_i$, highlighted in~\cref{fig:black_body} (same color code), i.e. $\nu_c=\nu_i$ and $L=c/(2\nu_i)$. The diagonal line represents the free-space density of states.}
    \label{fig:log_vs_om}
\end{figure*}
\indent \cref{fig:log_vs_om} displays the PLDOS as a function of frequency, for a dipole oriented parallel (top row) and perpendicular (bottom row) to the mirror planes, i.e. with the dipole lying  in the plane of the mirrors and along the cavity axis, respectively. The diagonal line corresponds to the free-space PLDOS $\rho_{\mathrm{free}}(\omega)$. It is clear that at low frequencies, $\nu \ll \nu_{c}$ (near-field regime), the total PLDOS inside the cavity~[\cref{fig:log_vs_om}a,d] can be significantly enhanced with respect to its free-space value. As shown in~\cref{fig:log_vs_om}c, f, this enhancement is driven by the presence of evanescent modes, which increase the total PLDOS by several orders of magnitude with respect to the free-space density. As an illustrative example, consider the cavity tuned to the peak of the blackbody energy density at room temperature corresponding to $L_3= 8.50\, \mathrm{\mu}$m ($\nu_3=17.6$ THz). The PLDOS exceeds the free-space density for frequencies $\nu \lesssim 100$ GHz for a parallel dipole orientation and up to $\nu \lesssim 10$ THz for a perpendicular dipole orientation. As the cavity length decreases, the frequency range over which this enhancement occurs shifts towards higher frequencies. A further feature emerging from the evanescent contribution is the appearance of a pronounced peak, which shifts towards the surface plasma frequency $\nu_{sp} = \nu_p/\sqrt{2} =\omega_p/(2\pi\sqrt{2}) \simeq 1924$ THz as the cavity length decreases. This reflects the increasingly efficient coupling to the surface plasmon-polariton resonance supported at the gold-vacuum interface~\cite{Raether, Pitarke}. For sufficiently small cavity lengths, the contribution of the surface plasmon-polariton mode to the evanescent PLDOS becomes large enough such that the PLDOS exceeds again the free-space density of states. A more detailed analysis of this behavior is presented in Fig. 1 of the Supplemental Material, where cavities nearly in resonance with $\nu_{sp}$ are considered. Nevertheless, note that the surface plasmon-polariton is not expected to contribute to heat transfer since it is too high in energy. At higher frequencies, far from the cavity resonance (i.e. $\nu \gg \nu_c$ or $\lambda \ll L$), the evanescent contribution to the PLDOS vanishes, and the propagating modes dominate. In this regime, the propagating contribution asymptotically approaches the free-space density of states, indicating that the influence of the cavity boundaries progressively diminishes, as expected. \\
\begin{figure*}[t]
    \centering
    \includegraphics[width=1.0\textwidth]{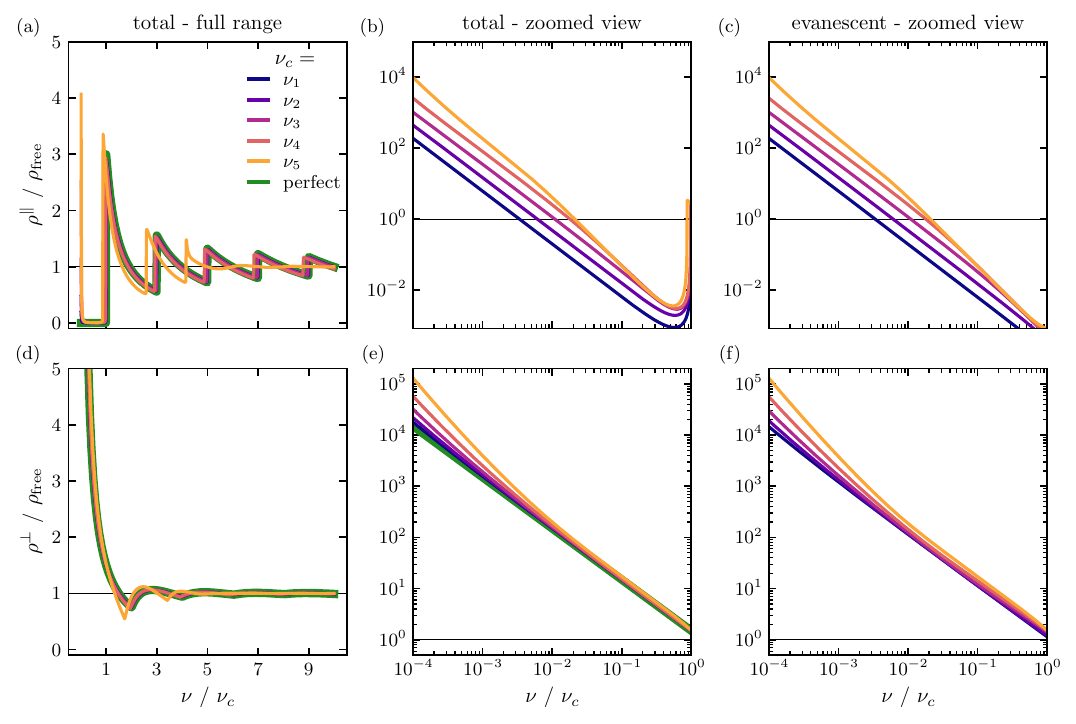}
    \caption{PLDOS vs ratio between frequency and cavity fundamental frequency for different dipole orientations. (a), (d) Total, (b), (e) Total (zoomed view) and (c), (f) evanescent (zoomed view) PLDOS normalized to the free-space for a parallel (upper row) and perpendicular (lower row) dipole orientation, evaluated at the center of the cavity, as functions of the ratio between frequency and cavity fundamental frequency, where $\rho^{\parallel/\perp}/\rho_{\mathrm{free}} = \rho(\hat{\bm{e}}_{x/z}, z = L/2, 2 \pi \nu)/\rho_{\mathrm{free}}(2 \pi \nu)$. The five curves correspond to five different cavities with a cavity fundamental frequency $\nu_c$ matching the five frequencies $\nu_i$,  highlighted in~\cref{fig:black_body} (same color code), i.e. $\nu_c=\nu_i$ and $L=c/(2\nu_i)$.}
    \label{fig:lin_vs_omc}
\end{figure*}
\indent To better understand the trends of the PLDOS in terms of $\nu/\nu_c$ at different cavity lengths and to compare to the ideal-mirror case, ~\cref{fig:lin_vs_omc} shows the total and evanescent parts of the PLDOS normalized to the free-space $\rho_{\mathrm{free}}(\omega)$ and as a function of the normalized frequency $\nu/\nu_c$.~\cref{fig:lin_vs_omc}a,d show that, for $\nu > \nu_c$, the total density of states closely matches the results for perfect mirrors~\cite{Fassioli}. The main deviation from the perfect mirror case occurs for the smallest cavity, i.e. the one corresponding to $\nu_c = \nu_5$ ($634.60$ THz). The discrepancy arises because, as explained before, the coupling to the surface plasmon-polariton mode starts to become more evident as the cavity length decreases. To better resolve the low-frequency behavior, we also present the density of states over a restricted range $\nu \ll \nu_c$, where the evanescent contribution is expected to dominate, in ~\cref{fig:lin_vs_omc}b, c, e, f. For both in-plane and out-of-plane dipole orientations, evanescent modes lead to a substantial enhancement of the density of states with respect to both the free-space and the perfect mirror cases at sufficiently low frequencies.

\subsection{PLDOS vs cavity length}

\begin{figure*}[t]
    \centering
    \includegraphics[width=1.0\textwidth]{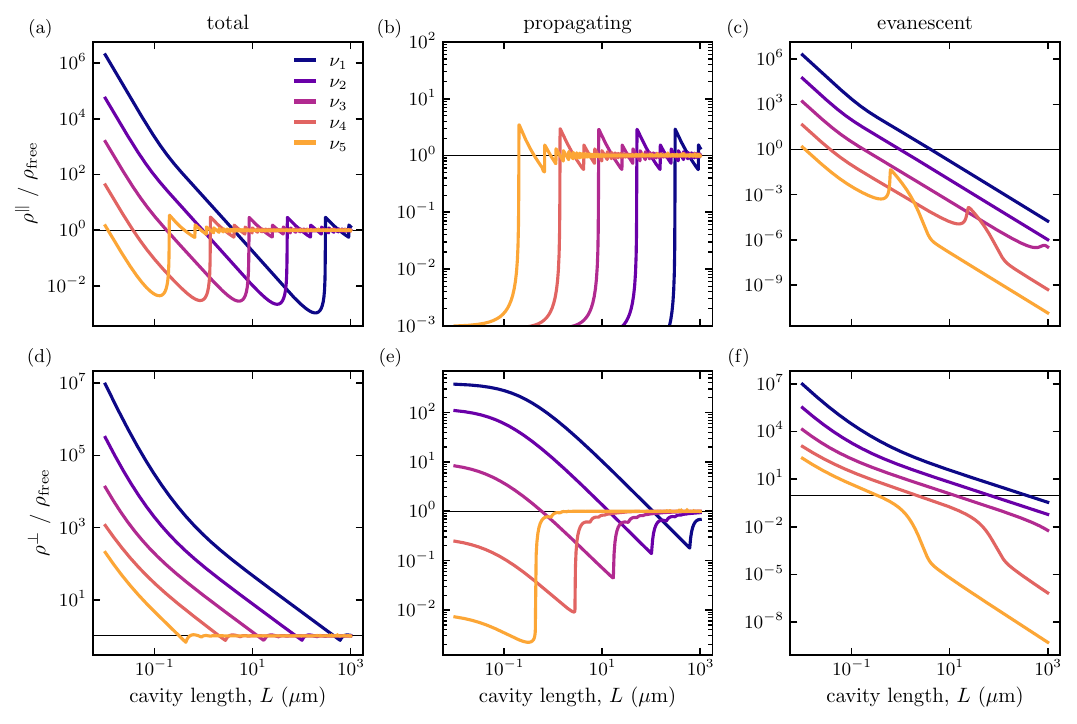}
    \caption{PLDOS vs cavity length for different dipole orientations. (a), (d) Total, (b), (e) propagating and (c), (f) evanescent PLDOS normalized to free-space for a parallel (upper row) and a perpendicular (lower row) dipole orientation, evaluated at the center of the cavity, as functions of cavity length L, computed for the five different frequencies $\nu_i$ highlighted in~\cref{fig:black_body} (same color code). Here $\rho^{\parallel/\perp}/\rho_{\mathrm{free}} = \rho(\hat{\bm{e}}_{x/z}, z = L/2, 2 \pi \nu_i)/\rho_{\mathrm{free}}(2 \pi \nu_i)$.  }
    \label{fig:log_vs_length}
\end{figure*}

\cref{fig:log_vs_length} presents the PLDOS for a given frequency, as a function of the cavity length, with the frequency fixed to the five representative values $\nu_i$ in~\cref{fig:black_body}. It is evident that reducing the cavity length leads to a pronounced increase in the evanescent contribution to the PLDOS. For example, at the frequency corresponding to the peak of the blackbody energy density spectrum at room temperature ($\nu_3=$ 17.6 THz), the evanescent PLDOS exceeds the free-space density for cavity lengths $L \lesssim 100$ nm in the case of a parallel dipole orientation, and for $L \lesssim 10\, \mu$m for a perpendicular dipole orientation, and therefore one can also expect an enhancement in the radiative heat transfer at these cavity lengths. Conversely, as the cavity length increases, electromagnetic confinement weakens, the evanescent contribution vanishes, and the propagating contribution approaches the free-space density of states, as expected.

\section{Near-field radiative heat transfer}
\label{sec:RHT}

\begin{figure}
    \centering
    \includegraphics[width=\columnwidth]{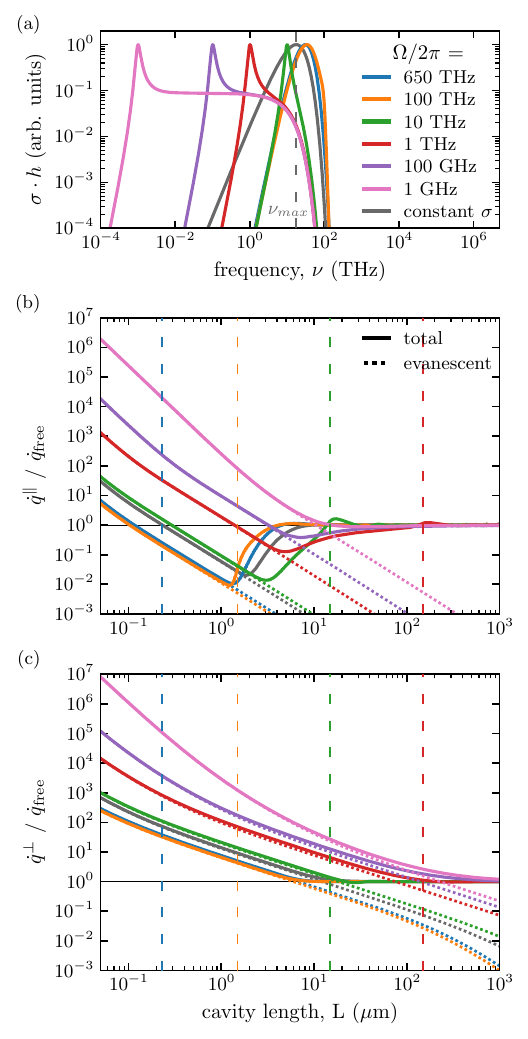}
    \caption{Heat flow between a Lorentz absorber at $T_m \to 0$ K and cavity mirrors at $T_{\mathrm{gold}} = 300$ K for different values of the absorber central frequency $\Omega$ ($\gamma = 0.3\,\Omega$), and for a frequency-independent conductivity $\sigma=constant$. (a) Spectral density weighted by conductivity, $h(\omega,T_{gold},T_m) \times \sigma(\omega)$, in~\cref{eq:heat_flow}. (b), (c) Heat flow per volume normalized to the free-space value for parallel (b) and perpendicular (c) dipole orientations (evanescent contributions are shown as dotted lines). Vertical dashed lines show the corresponding cavity length for a cavity tuned to each $\Omega$, i.e. $L_c=\pi c/\Omega$ (for $\Omega<100$ GHz, $L_c>1.5$ mm is out of range). }
    \label{fig:heatflow}
\end{figure}

We now turn to the analysis of the radiative heat transfer,  with the goal of identifying the regime in which near-field contributions become significant. Heat flow per unit volume is calculated according to~\cref{eq:heat_flow}. The gold mirrors are assumed to be in thermal equilibrium with the surrounding photonic bath, such that both are maintained at the same temperature $T_{\mathrm{gold}}$. Accordingly, $T_l=T_{\mathrm{gold}}$ denotes the temperature of the electromagnetic environment in both the free-space and in-cavity calculations. In the following, we fix $T_{\mathrm{gold}} = 300$ K, while the material temperature is set to $T_m \to 0$ K. To model the material conductivity, we consider a single Lorentz oscillator centered at a frequency $\Omega$, with optical conductivity
\begin{equation}
\label{sigma_w}
    \sigma(\omega) = \epsilon_0 \frac{\omega_P^2\, \gamma\, \omega^2 }{(\Omega^2 - \omega^2)^2 + \omega^2 \, \gamma^2}\,, \quad \gamma = 0.3\, \Omega\,,
\end{equation}
as in~\cite{Fassioli}. 
It is important to note that  at frequencies exceeding the blackbody energy density peak, $\nu \gg \nu_{max} \approx 2.82 k_B T_l/(2 \pi \hbar)$ [\cref{fig:black_body}], the contribution of the optical conductivity to heat flow in~\cref{eq:heat_flow} becomes exponentially suppressed  by the tail of the Bose-Einstein distribution. This behavior is illustrated in~\cref{fig:heatflow}a which shows the optical conductivity $\sigma(\omega)$ weighted by the energy density $h(\omega, T_l,T_m)$ for several values of $\Omega$, together with the case of  a flat conductivity. At the same time, irrespective of the oscillator frequency and even when $\Omega/(2\pi) \ll \nu_{max}$, modes with frequencies near $\nu_{max}$ are expected to be relevant and may even dominate heat transfer because the free-space PLDOS contained in $h(\omega, T_l,T_m)$ scales as $\propto \omega^2$.  Nevertheless,  lowering $\Omega$ shifts the spectral peak to lower frequencies. Consequently, one can expect that lowering the oscillator frequency allows for a modification of heat flow with longer cavities.\\
\indent ~\cref{fig:heatflow}b, c  show the heat flow normalized to the corresponding free-space value $\dot{q}_{\mathrm{free}}$ as a function of cavity length for several values of $\Omega$ for both dipole orientations. As shown, cavities ranging from nanometer to micrometer scales can enhance heat transfer by several orders of magnitude for both in-plane and out-of-plane dipole orientations, with the out-of-plane orientation producing the larger enhancement. As anticipated, the effect is also sensitive to the material absorption band: lowering the absorber central frequency allows heat-flow enhancement with longer cavity lengths. For example, for $\Omega/(2 \pi) = 10 $ THz, evanescent modes significantly enhance ($> 50$ times) heat flow  in cavity lengths below $L \lesssim 100$ nm for parallel dipole orientation, and below cavity lengths of $L \lesssim  1\, \mu$m for a perpendicular dipole orientation. Similar enhancements appear in longer cavities for $\Omega/(2 \pi) = 100$ GHz ($L \lesssim 1\, \mu$m for a parallel dipole, and  $L \lesssim  10\, \mu$m for a perpendicular dipole orientation). \\
\indent The present results are in stark contrast to those in our previous work considering ideal cavity mirrors (see~\cite{Fassioli} and Fig. 2 of Supplemental Material). 
For \textit{in-plane dipole orientations}, a perfect mirror can enhance the PLDOS near the cavity resonances, with a maximum threefold enhancement at $\nu=\nu_c$, but completely suppresses the PLDOS for $\nu<\nu_c$. Consequently, we  found that only cavities tuned to the frequency range where free-space heat flow dominates, namely the $\sim$1-10 THz range at room temperature, could modestly enhance heat flow, owing to the small increase in the PLDOS near $\nu_c$. The enhancement was maximized ($\sim$ 1.5) when both the cavity resonance and the absorber central frequency were tuned to $\nu_{max}$. 
For longer cavities, $\nu_c\ll \nu_{max}$, the enhancement was negligible because the cavity does not significantly modify the PLDOS in the mid-IR regime. 
Conversely, for shorter cavities,  with $\nu_c\gg \nu_{max}$, heat flow was strongly suppressed  due to the suppression of the PLDOS in the mid-IR irrespective of absorber frequency. Similarly, heat flow was suppressed for low-frequency absorbers satisfying $\Omega/(2\pi)\ll \nu_c$, i.e., for cavity lengths sufficiently short that the cavity resonance lies well above the absorber frequency. In contrast, in the present study, we find that cavities with $\nu_c\gg\Omega/(2\pi)$ can significantly enhance the PLDOS near $\Omega$, rather than suppress it as in the case of ideal mirrors. Moreover, cavities tuned well above the mid-IR, $\nu_c\gg\nu_{max}$, corresponding to cavity lengths in the nanometric range, can strongly enhance rather than suppress the PLDOS at the mid-IR. This provides a mechanism for enhancing heat transfer to an in-plane absorber centered not only in the mid-IR but also at lower frequencies. Importantly, the resulting enhancement can be several orders of magnitude larger than the relatively modest enhancement achievable with perfect mirrors.\\
\indent For \textit {out-of-plane dipole orientations}, ideal cavity mirrors increase the PLDOS with respect to free space below $\nu_c$. This implies that heat transfer can  also be significantly enhanced in ideal cavities as long as $\nu_c$ is large (cavity length is short) enough  to enhance PLDOS in the regime where heat flow dominates in free space (see Fig. 2 of Supplemental Material and~\cite{Fassioli}). Nevertheless, the enhancement achieved with gold mirrors can be a few orders of magnitude larger than with ideal mirrors. Notably, for out-of-plane dipole orientations and high frequency absorbers ($\Omega / (2 \pi) \gtrsim 100$ THz), cavities resonant to the oscillator frequency, i.e. $\nu_c \sim \Omega/(2\pi)$, can enhance heat transfer by a factor of $\sim$ 10-100, which is interesting since many experiments aim at placing molecules or materials in cavities tuned to a specific material optical transition. 

\section{Discussion and conclusions}
\label{sec:conclusions}
The previous results demonstrate that near-field evanescent modes can strongly enhance the photonic local density of states and, as a consequence, the radiative heat flow in nanometer to micrometer metallic imperfect cavities. The effect is sensitive to the dipole orientation, and the spectral dependence of the optical conductivity plays a crucial role in determining the magnitude and spatial range of heat-transfer enhancement.
As the absorber frequency is lowered, near-field effects appear at larger cavity separations. In general, heat flow is increased by cavities whose fundamental frequency is much higher than the oscillator frequency ($\nu_c \gg \Omega/(2 \pi)$), and therefore no near-field effects are expected in cavities tuned to the absorber central frequency. The only exception occurs when the material oscillator frequency is higher than the frequency at which the blackbody spectral energy density reaches its maximum ($\Omega \gtrsim 2\pi\nu_{max}$) and the dipole is oriented perpendicular to the mirrors, in which case radiative heat flow can be significantly increased even in resonant cavities, i.e. for $\nu_c\sim \Omega/(2\pi)$. \\
\indent The contribution of evanescent modes to the PLDOS in nanometer and up to micrometer cavities raises the possibility that near-field effects may alter the heat load on a range of absorbers used in polariton chemistry experiments, even those resonant with the cavity fundamental frequency. Electronic strong-coupling experiments (molecular excitons in J-aggregates, dyes, etc.) generally employ planar Fabry–Pérot cavities with mirror separations of a few hundred nanometers~\cite{DelPo2020}, within the range where our results predict significant near-field enhancement. Moreover, the PLDOS available to lower energy molecular excitations, and in particular vibrations, will be greatly enhanced in such cavities and could, in principle, alter vibrationally mediated processes. Vibrational strong coupling experiments, on the other hand, use cavity separations in the 1-100 $\mu$m range, where our results also show an enhancement in heat flow, in particular for perpendicular dipole orientations~\cite{Hirai, Grafton, Ebbesen}. Overall we expect cavity-modified heat load to dominate at frequencies $\nu \sim \nu_c$ or $\ll \nu_c$ for significant near-field effects, which is consistent with polariton chemistry experiments showing modified rates at normal incidence. However, our mechanism does not require the achievement of strong coupling as most experiments seem to invoke. A more in-depth analysis of the relevance of near-field effects in cavity-modified molecular processes is left for future work.\\
\indent In this work we studied the heat flow between the cavity mirrors and a material with purely electric polarizability. However, in materials with a magnetic polarizability, the cavity also modifies heat flow through the changes in the magnetic PLDOS. Studies considering near-field heat transfer between a metallic nanoparticle and a semi-infinite mirror~\cite{Chapuis_nanoparticles} or between two parallel metallic plates~\cite{Chapuis}, show that the magnetic near-field contribution can even dominate heat transfer at very short distances. It will be interesting to explore effects similar to those studied in this work in materials with a magnetic polarizability.\\
\indent To summarize, in this work, we have systematically analyzed the contribution of evanescent modes to the PLDOS in a planar cavity formed by two semi-infinite gold mirrors and their role in enhancing radiative heat flow for a material embedded inside the cavity, extending the framework of~\cite{Fassioli} to the case of a realistic, dissipative metallic boundary. By employing the electric dyadic Green's function formalism~\cite{Barnes}, we have obtained propagating and evanescent contributions to the PLDOS for both in-plane and out-of-plane dipole orientations. Unlike a normal-mode expansion, this approach remains applicable in the presence of material dissipation, where the system eigenfrequencies become complex. Evanescent modes substantially increase the PLDOS at frequencies well below the cavity fundamental frequency, $\nu \ll \nu_c$, i.e., far from the cavity resonance. Such enhancement at subwavelength separations is consistent with past works analyzing PLDOS near a single mirror ~\cite{Carminati, Joulain, Joulain_2003, Carminati_2006, Castanie, Shchegrov, Carminati_1999}. Since, under our model assumptions, heat transfer is directly related to the PLDOS, the enhancement of the PLDOS inside the cavity due to evanescent fields has direct consequences on heat transfer. In particular, we found that evanescent cavity modes yield a significant modification of the heat exchange between a material and the gold mirrors, irrespective of the oscillator frequency. This is in contrast to our previous findings for an ideal cavity formed by perfect mirrors~\cite{Fassioli},  especially for in-plane dipoles, where an appreciable though modest heat load modification required the cavity and the oscillator to be tuned near the mid-IR frequency range.\\
\indent The framework developed here provides a systematic way to evaluate the relative weight of far-field and near-field effects to the radiative heat transfer and is directly applicable to experimentally relevant platforms in which nanometric and micrometric metallic cavities are employed, such as Fabry-Pérot cavities used in polaritonic chemistry~\cite{Mandal, Yuen}. A quantitative account of near-field terms may prove relevant for a proper energy balance in such systems~\cite{Fassioli_2021, Li}.\\
\indent A further direction for future work is to consider alternative materials for the cavity mirrors, which can have a lower plasma frequency and therefore shift the plasmonic contribution to the near-field PLDOS towards the spectral range relevant for heat transfer at room temperature. A potentially interesting material class in this respect could be narrow-gap semiconductors, which have low carrier densities and small effective masses and allow strongly confined plasmonic excitations at much lower frequencies than conventional metals. For example, strongly confined THz surface-plasmon cavities based on InSb have recently been realized experimentally~\cite{Aupiais}. Although this work does not address radiative heat transfer, it shows that, for sufficiently strong plasmonic confinement, the response of InSb departs from a local electrodynamic description, with signatures of nonlocality and Landau damping emerging already at THz frequencies. Nonlocal effects may therefore become relevant at substantially larger length scales in such materials than in Au. Theoretical calculations of near-field radiative heat transfer between doped semiconductors have indeed shown that spatial dispersion can regularize the large-$k_{\parallel}$ response and lead to a saturation of the heat-transfer coefficient at short distances~\cite{Singer}, and experimentally, nonlocal effects in near-field heat transfer have also been reported in a system involving an ultrathin plasmonic layer~\cite{Salihoglu}. \\
\indent Another direction would be to consider an atomically thin material whose position within the cavity can be controlled. In the present work, the absorber is placed at the center of the cavity, whereas the evanescent contribution is intrinsically sensitive to its distance from the interfaces. For a two-dimensional material, its position along the cavity axis would therefore provide an additional control parameter: shifting the layer towards one mirror could continuously tune the relative contributions of propagating and evanescent modes to the radiative heat exchange.\\
\indent A natural extension for future work is the inclusion of non-equilibrium effects. In the present work, the material and the photon bath, weakly interacting, are assumed to be in thermal equilibrium at well-defined temperatures. However, more general situations may arise when the material interacts simultaneously with different photonic environments or light-matter coupling is strong, leading to steady states that cannot be described by thermal equilibrium distributions. Recent theoretical progress in this direction has been reported in~\cite{Flores-Calderon}, where the authors consider a quantum material coupled to two distinct photon baths and show that electrons and cavity photons can be driven away from thermal equilibrium even in the steady state. This breakdown of thermal equilibrium leads to a reduced symmetry of the steady-state electronic distribution function and to qualitative modifications of material properties. Incorporating these effects in realistic dissipative cavities supporting near-field electromagnetic modes could provide further insight into cavity-controlled thermal and electronic phenomena beyond equilibrium.

\begin{acknowledgments}
    This work was supported by the Gordon and Betty Moore Foundation
through the grant CENTQC (No. GBMF12213). M.E. acknowledges funding from the Deutsche Forschungsgemeinschaft (DFG) via the Cluster of Excellence CUI: Advanced Imaging of Matter -- EXC 2056 -- project ID 390715994. Funding is acknowledged through the Deutsche Forschungsgemeinschaft through OPTIMAL-FOR5750 - 531215165 (Project P1 and P8).

\end{acknowledgments}

\newpage

\bibliography{MyLibrary}

\end{document}

\preprint{}

\title{\textbf{Supplemental material for "Near-field contributions to cavity-mediated radiative heat transfer"}}

\author{Antonio Mastropasqua}
\affiliation{Department of Physics, Friedrich-Alexander-Universität Erlangen-Nürnberg, 91058 Erlangen, Germany}

\author{Martin Eckstein}
\affiliation{Department of Physics, Universität Hamburg, 22607 Hamburg, Germany}
\affiliation{The Hamburg Centre for Ultrafast Imaging, Hamburg, Germany}

\author{Daniele Fausti}
\affiliation{Department of Physics, Friedrich-Alexander-Universität Erlangen-Nürnberg, 91058 Erlangen, Germany}
\affiliation{Department of Physics, Università degli Studi di Trieste, 34127 Trieste, Italy}

\author{Francesca Fassioli}
\affiliation{Department of Physics, Friedrich-Alexander-Universität Erlangen-Nürnberg, 91058 Erlangen, Germany}

\maketitle

\section{Explicit derivation of propagating and evanescent contributions to the partial local density of states}
In this section, we provide the details of the calculation of the electric PLDOS for a planar Fabry-Pérot cavity bounded by two dissipative gold mirrors. The derivation follows the standard dyadic Green's function formalism for planar multilayer structures, as established in the references cited in the main text~\cite{Chew, Barnes, Joulain_2003}. The purpose of this derivation is not to introduce a new expression for the PLDOS, but rather to apply the established formalism to the present cavity geometry and, in particular, to explicitly separate the propagating and evanescent contributions. Starting from the dyadic Green's function of the cavity, we first obtain the total electric PLDOS and then decompose the corresponding wavevector integrals into the propagating ($k_{\parallel} \leq k_0$) and evanescent ($k_{\parallel} > k_0$) sectors. The resulting expressions are those used in the main text to analyze the relative contributions of propagating and evanescent modes to the PLDOS and radiative heat exchange.\\
\indent We consider a cavity formed by two semi-infinite gold layers at $z=0$ and $z=L$, so that $L$ is the length of the cavity. Both exterior media have the same Drude dielectric response $\epsilon^{\mathrm{Au}}(\omega)$. The Fresnel reflection coefficients for the transverse electric (TE, $r_1$) and transverse magnetic (TM, $r_2$) polarizations at the vacuum-gold interface are given by
\begin{subequations}
    \begin{align}
    r_1(\omega,k_{\parallel}) &= \frac{\sqrt{k_0^2 - k_{\parallel}^2} - \sqrt{\epsilon^{\mathrm{Au}}(\omega) k_0^2 - k_{\parallel}^2}}{\sqrt{k_0^2 - k_{\parallel}^2} + \sqrt{\epsilon^{\mathrm{Au}}(\omega) k_0^2 - k_{\parallel}^2}}\,, \\
    r_2(\omega,k_{\parallel}) &= \frac{\epsilon^{\mathrm{Au}}(\omega) \sqrt{k_0^2 - k_{\parallel}^2} - \sqrt{\epsilon^{\mathrm{Au}}(\omega) k_0^2 - k_{\parallel}^2}}{\epsilon^{\mathrm{Au}}(\omega) \sqrt{k_0^2 - k_{\parallel}^2} + \sqrt{\epsilon^{\mathrm{Au}}(\omega) k_0^2 - k_{\parallel}^2}}\,,
    \end{align}
\end{subequations}
where $k_0 = \omega/c$ is the free-space wavevector, $k_{\parallel}$ is its in-plane component and $k_z = \sqrt{k_0^2 - k_{\parallel}^2}$ is its out-of-plane component.
The dyadic Green's function for this cavity geometry can be written as the sum of the free-space contribution and a term accounting for the reflections from the two cavity mirrors. The free-space contribution is given by
\begin{equation}
    \mathbb{G}_0(\bm{r},\bm{r}',\omega) = \sum\limits_{\zeta = 1,2} \int  \frac{d^2k_{\parallel}}{(2\pi)^2} \frac{i}{2 k_z} \hat{\bm{e}}_{\zeta}^{(-)} \otimes \hat{\bm{e}}_{\zeta}^{(-)} e^{i k_z (z'-z)} e^{i \bm{k}_{\parallel} \cdot (\bm{r}_{\parallel} - \bm{r}'_{\parallel})}\,, \quad \text{for}\,  z - z'<0\,,
\end{equation}
where the unit vectors corresponding to the two polarizations are
\begin{subequations}
    \begin{align}
        \hat{\bm{e}}_1^{(\pm)} &= \frac{1}{k_{\parallel}} \left(k_x \hat{\bm{e}}_y - k_y \hat{\bm{e}}_x \right)\,, \\
        \hat{\bm{e}}_2^{(\pm)} &= \frac{1}{k_0} ( k_{\parallel} \hat{\bm{e}}_z \mp k_z \hat{\bm{e}}_{k \parallel} )\,.
    \end{align}
\end{subequations}
The contribution arising from reflections at the two mirrors contains an infinite series of multiple reflections. A wave emitted toward the mirror at $z=0$ undergoes the following sequence of reflections
\begin{equation}
\begin{split}
    &\hat{\bm{e}}_{\zeta}^{(-)} e^{-ik_z(z-z')} + \hat{\bm{e}}_{\zeta}^{(+)} r_{\zeta}\, e^{ik_z(z+z')} + \hat{\bm{e}}_{\zeta}^{(-)} r_{\zeta}^2\, e^{ik_z(2L-z+z')} + \hat{\bm{e}}_{\zeta}^{(+)} r_{\zeta}^3\, e^{ik_z(2L+z+z')} + \ldots \\
    & =\hat{\bm{e}}_{\zeta}^{(-)} e^{-ik_z(z-z')} + \frac{ \hat{\bm{e}}_{\zeta}^{(+)} r_{\zeta}\, e^{ik_z(z+z')}}{1- r_{\zeta}^2\, e^{2 i k_z L}} + \frac{ \hat{\bm{e}}_{\zeta}^{(-)} r_{\zeta}^2\, e^{ik_z(2L-z+z')}}{1- r_{\zeta}^2\, e^{2 i k_z L}}\,,
\end{split}
\end{equation}
while a wave emitted toward the mirror at $z=L$ undergoes the sequence
\begin{equation}
\begin{split}
    &\hat{\bm{e}}_{\zeta}^{(+)} e^{ik_z(z-z')} + \hat{\bm{e}}_{\zeta}^{(-)} r_{\zeta}\, e^{ik_z(2L-z-z')} + \hat{\bm{e}}_{\zeta}^{(+)} r_{\zeta}^2\, e^{ik_z(2L+z-z')} + \hat{\bm{e}}_{\zeta}^{(-)} r_{\zeta}^3\, e^{ik_z(4L-z-z')} + \ldots \\
    & =\hat{\bm{e}}_{\zeta}^{(+)} e^{ik_z(z-z')} + \frac{ \hat{\bm{e}}_{\zeta}^{(-)} r_{\zeta}\, e^{ik_z(2L-z-z')}}{1- r_{\zeta}^2\, e^{2 i k_z L}} + \frac{ \hat{\bm{e}}_{\zeta}^{(+)} r_{\zeta}^2\, e^{ik_z(2L+z-z')}}{1- r_{\zeta}^2\, e^{2 i k_z L}}\,.
\end{split}
\end{equation}
The total dyadic Green's function can therefore be written as
\begin{equation}
\begin{split}
    \mathbb{G}(\bm{r},\bm{r}',\omega) &= \mathbb{G}_0(\bm{r},\bm{r}',\omega) \\
    &\ \ \ +\sum\limits_{\zeta = 1,2} \int  \frac{d^2k_{\parallel}}{(2\pi)^2} \frac{i}{2 k_z} \left[\frac{ \hat{\bm{e}}_{\zeta}^{(+)} \otimes \hat{\bm{e}}_{\zeta}^{(-)} r_{\zeta}\, e^{ik_z(z+z')}}{1- r_{\zeta}^2\, e^{2 i k_z L}} + \frac{ \hat{\bm{e}}_{\zeta}^{(-)} \otimes \hat{\bm{e}}_{\zeta}^{(-)} r_{\zeta}^2\, e^{ik_z(2L-z+z')}}{1- r_{\zeta}^2\, e^{2 i k_z L}} \right] e^{i \bm{k}_{\parallel} \cdot (\bm{r}_{\parallel} - \bm{r}'_{ \parallel})} \\
    &\ \ \ + \sum\limits_{\zeta = 1,2} \int  \frac{d^2k_{\parallel}}{(2\pi)^2} \frac{i}{2 k_z} \left[\frac{ \hat{\bm{e}}_{\zeta}^{(-)} \otimes \hat{\bm{e}}_{\zeta}^{(+)} r_{\zeta}\, e^{ik_z(2L-z-z')}}{1- r_{\zeta}^2\, e^{2 i k_z L}} + \frac{ \hat{\bm{e}}_{\zeta}^{(+)} \otimes \hat{\bm{e}}_{\zeta}^{(+)} r_{\zeta}^2\, e^{ik_z(2L+z-z')}}{1- r_{\zeta}^2\, e^{2 i k_z L}}\right] e^{i \bm{k}_{\parallel} \cdot (\bm{r}_{\parallel} - \bm{r}'_{ \parallel})}\,.
\end{split}
\end{equation}
Substituting the Green's function into the definition of the PLDOS gives
\begin{equation}
\begin{split}
    \rho(\hat{\bm{e}}_a, \bm{r},\omega) &= \frac{\omega^2}{3 \pi^2 c^3} \\
    &\ \ \ + \frac{\omega}{\pi c^2} \text{Re}\left\{ \sum\limits_{\zeta = 1,2} \int  \frac{d^2k_{\parallel}}{(2\pi)^2} \frac{1}{k_z} \left[\frac{ (\hat{\bm{e}}_a \cdot \hat{\bm{e}}_{\zeta}^{(+)}) (\hat{\bm{e}}_{\zeta}^{(-)} \cdot \hat{\bm{e}}_a) r_{\zeta}\, e^{2i k_z z}}{1- r_{\zeta}^2\, e^{2 i k_z L}} + \frac{ (\hat{\bm{e}}_a \cdot \hat{\bm{e}}_{\zeta}^{(-)})^2 r_{\zeta}^2\, e^{2i k_z L}}{1- r_{\zeta}^2\, e^{2 i k_z L}} \right] \right\} \\
    &\ \ \ + \frac{\omega}{\pi c^2} \text{Re}\left\{ \sum\limits_{\zeta = 1,2} \int  \frac{d^2k_{\parallel}}{(2\pi)^2} \frac{1}{k_z} \left[\frac{ (\hat{\bm{e}}_a \cdot \hat{\bm{e}}_{\zeta}^{(-)}) (\hat{\bm{e}}_{\zeta}^{(+)} \cdot \hat{\bm{e}}_a) r_{\zeta}\, e^{2i k_z (L-z)}}{1- r_{\zeta}^2\, e^{2 i k_z L}} + \frac{ (\hat{\bm{e}}_a \cdot \hat{\bm{e}}_{\zeta}^{(+)})^2 r_{\zeta}^2\, e^{2i k_z L}}{1- r_{\zeta}^2\, e^{2 i k_z L}} \right] \right\}\,.
\end{split}
\end{equation}
For dipoles oriented parallel ($\hat{\bm{e}}_a = \hat{\bm{e}}_x$) and perpendicular ($\hat{\bm{e}}_a = \hat{\bm{e}}_z$) to the plane of the cavity mirrors, this expression yields the corresponding PLDOS. Normalizing by the free-space PLDOS $\rho_{\mathrm{free}}(\omega) = \omega^2 / (3 \pi^2 c^3)$, we obtain
\begin{subequations}
    \begin{align}
        \frac{\rho^{\parallel}(\hat{\bm{e}}_x, \bm{r}, \omega)}{\rho_{\mathrm{free}}(\omega)} = 1  &+ \frac{3}{4 k_0} \text{Re}\left\{\int_0^{\infty} dk_{\parallel} \frac{k_{\parallel}}{k_z} \left[ \frac{r_1 (e^{2 i k_z z} + e^{2 i k_z(L-z)})+2 r_1^2 e^{2ik_z L}}{1-r_1^2 e^{2i k_z L}} \right] \right\} \nonumber \\
        &+ \frac{3}{4 k_0} \text{Re}\left\{\int_0^{\infty} dk_{\parallel} \frac{k_{\parallel} k_z}{k_0^2} \left[ \frac{-r_2 (e^{2 i k_z z} + e^{2 i k_z(L-z)})+2 r_2^2 e^{2ik_z L}}{1-r_2^2 e^{2i k_z L}} \right] \right\}\,, \label{eq:par_rho} \\
        \frac{\rho^{\perp}(\hat{\bm{e}}_z, \bm{r}, \omega)}{\rho_{\mathrm{free}}(\omega)} = 1  &+ \frac{3}{2 k_0} \text{Re}\left\{\int_0^{\infty} dk_{\parallel} \frac{k_{\parallel}^3}{k_z k_0^2} \left[ \frac{r_2 (e^{2 i k_z z} + e^{2 i k_z(L-z)})+2 r_2^2 e^{2ik_z L}}{1-r_2^2 e^{2i k_z L}} \right] \right\}\,. \label{eq:perp_rho}
    \end{align}
\end{subequations}
\noindent ~\cref{eq:par_rho} and ~\cref{eq:perp_rho} are valid for an arbitrary position $z$ within the cavity. Computing these expressions for a dipole located at the center of the cavity ($z = L/2$), and splitting the wavevector integrals in the propagating ($k_{\parallel} \leq k_0$) and evanescent ($k_{\parallel} > k_0$) sectors, we finally derive the expressions present in the paper
\begin{subequations}\label{eq:rhoprop}
    \begin{align}
        \frac{\rho^{\parallel}_{\mathrm{prop}}(\hat{\bm{e}}_x, z = L/2, \omega)}{\rho_{\mathrm{free}}(\omega)} &= 1 + \frac{3}{2 k_0} \int_{0}^{k_0} dk_{\parallel} \frac{k_{\parallel}}{k_z} \frac{\text{Re} \left[r_1 e^{i k_z L}\right] - \left| r_1 \right|^2  }{\left| 1 - r_1 e^{i k_z L} \right|^2} \nonumber \\
        \ \ \ &- \frac{3}{2 k_0} \int_{0}^{k_0} dk_{\parallel} \frac{k_{\parallel} k_z}{k_0^2}  \frac{\text{Re} \left[r_2 e^{i k_z L}\right] + \left| r_2 \right|^2  }{\left| 1 + r_2  e^{i k_z L} \right|^2}\,, \label{eq:rhoprop1} \\
        \frac{\rho^{\perp}_{\mathrm{prop}}(\hat{\bm{e}}_z, z = L/2, \omega)}{\rho_{\mathrm{free}}(\omega)} &= 1 + \frac{3}{k_0} \int_{0}^{k_0} dk_{\parallel} \frac{k_{\parallel}^3}{k_0^2 k_z} \frac{\text{Re} \left[r_2 e^{i k_z L}\right] - \left| r_2 \right|^2  }{\left| 1 - r_2 e^{i k_z L} \right|^2}\,, \label{eq:rhoprop2}
        \end{align}
\end{subequations}
\begin{subequations}\label{eq:rhoevan}
    \begin{align}
        \frac{\rho^{\parallel}_{\mathrm{evan}}(\hat{\bm{e}}_x, z = L/2, \omega)}{\rho_{\mathrm{free}}(\omega)} &= \frac{3}{2 k_0} \int_{k_0}^{+\infty} dk_{\parallel} \frac{k_{\parallel}}{\mathcal{K}_z} e^{-\mathcal{K}_z L} \frac{\text{Im} \left[r_1 \right]}{\left| 1 - r_1 e^{- \mathcal{K}_z L} \right|^2} \nonumber \\
        \ \ \ &+ \frac{3}{2 k_0} \int_{k_0}^{+\infty} dk_{\parallel} \frac{k_{\parallel} \mathcal{K}_z}{k_0^2} e^{-\mathcal{K}_z L} \frac{\text{Im} \left[r_2\right]}{\left| 1 + r_2 e^{- \mathcal{K}_z L} \right|^2}\,, \label{eq:rhoevan1} \\
        \frac{\rho^{\perp}_{\mathrm{evan}}(\hat{\bm{e}}_z, z = L/2, \omega)}{\rho_{\mathrm{free}}(\omega)} &= \frac{3}{k_0} \int_{k_0}^{+\infty} dk_{\parallel} \frac{k_{\parallel}^3}{k_0^2 \mathcal{K}_z} e^{-\mathcal{K}_z L} \frac{\text{Im} \left[r_2 \right]}{\left| 1 - r_2  e^{- \mathcal{K}_z L} \right|^2}\,. \label{eq:rhoevan2}
    \end{align}
\end{subequations}

\section{PLDOS for a nanocavity with cavity frequency near the gold surface plasma frequency}
Given the local Drude dielectric response of gold
\begin{equation}
    \epsilon^{\mathrm{Au}}(\omega) = 1 - \frac{\omega_p^2}{\omega^2 + i \gamma \omega}
\end{equation}
the surface plasma frequency can be written in angular frequency as $\omega_{sp} = \omega_p / \sqrt{2}$ or, equivalently, as $\nu_{sp} = \omega_{sp}/(2\pi) = 1.924 \times 10^3$ THz. \\
In this section, we analyze the results for the PLDOS as a function of frequency, computed for a cavity length $L = 50$ nm. The corresponding cavity fundamental frequency $\nu_c = c/ (2L) = 2.998 \times 10^3$ THz lies in the vicinity of the gold surface plasma frequency.

\begin{figure}[h]
    \centering
    \includegraphics[width=1.\linewidth]{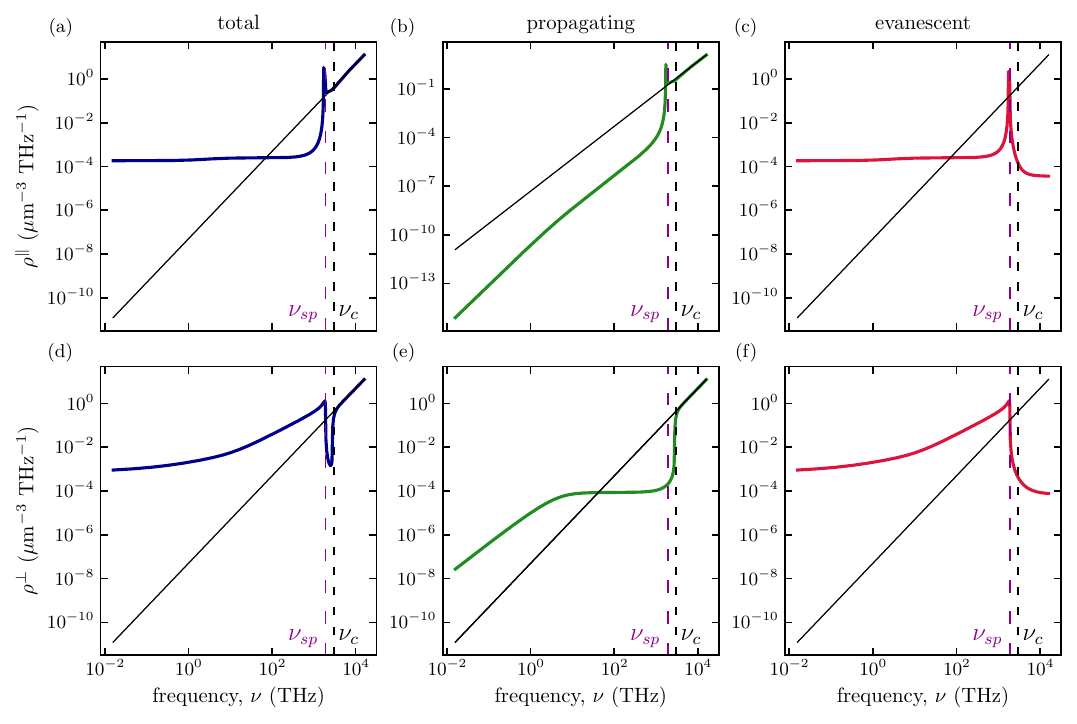}
    \caption{PLDOS vs frequency for different dipole orientations. (a), (d) Total, (b), (e) propagating and (c), (f) evanescent PLDOS for a parallel (upper row) and a perpendicular (lower row) dipole orientation, evaluated at the center of the cavity, as functions of frequency, where $\rho^{\parallel / \perp} = \rho(\hat{\bm{e}}_{x/z}, z = L/2, 2 \pi \nu)$. The densities are computed for a fixed cavity length $L= 50$ nm, corresponding to a cavity fundamental frequency $\nu_c = 2.998 \times 10^3$ THz. The solid diagonal line shows the free-space density of states. The vertical dashed lines indicate the surface plasma frequency $\nu_{sp}$ (purple) and the cavity fundamental frequency $\nu_c$ (black).}
    \label{fig:PLDOSvsom_small_cav}
\end{figure}
~\cref{fig:PLDOSvsom_small_cav} displays the PLDOS as a function of frequency, for a dipole oriented parallel (top row) and perpendicular (bottom row) to the mirror planes. As discussed in the main text, the evanescent PLDOS exceeds the free-space density at frequencies well below the cavity fundamental frequency. In addition, it exhibits a pronounced peak at the surface plasma frequency, near the cavity fundamental frequency, exceeding the free-space density of states. This enhancement originates from the coupling to the surface plasmon-polariton resonance supported by the gold-vacuum interface.

\section{Comparison with the heat flow in an ideal cavity}
In this section, we show the radiative heat flow for a cavity bounded by ideal mirrors, as previously investigated in~\cite{Fassioli}. Here we consider the same range of cavity lengths and oscillator frequencies, as in the main text, to allow for a direct comparison between ideal and gold mirrors. The same color code is used throughout the two sets of results.
\begin{figure}[h]
    \centering
    \includegraphics[width=1.\linewidth]{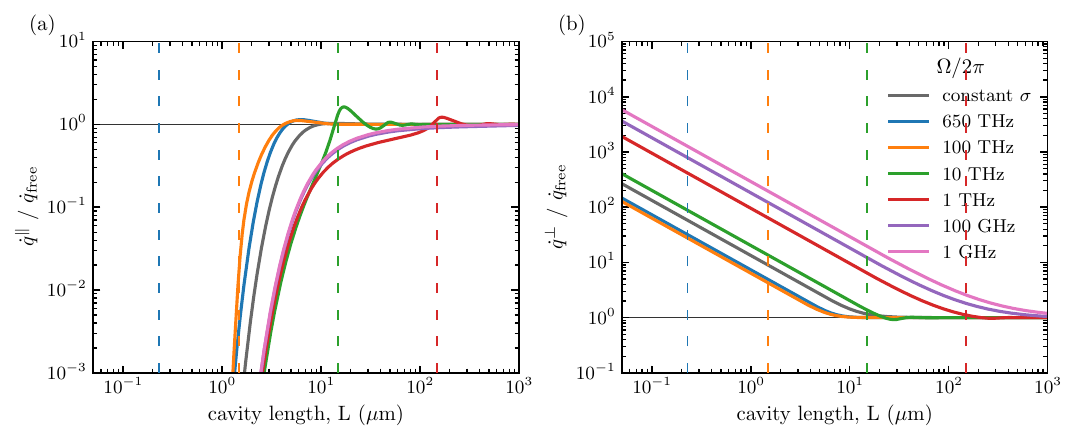}
    \caption{Heat flow between a Lorentz absorber at $T_m \to 0$ K placed in an ideal cavity and the external photon bath at temperature $T_{l} = 300$ K for different values of the absorber central frequency $\Omega$ ($\gamma = 0.3\,\Omega$), and for a frequency-independent conductivity $\sigma = constant$. Heat flow per volume normalized to the free-space value for parallel (a) and perpendicular (b) dipole orientations. Vertical dashed lines show the corresponding cavity length for a cavity tuned to each $\Omega$, i.e. $L_c=\pi c/\Omega$ (for $\Omega<100$ GHz, $L_c>1.5$ mm is out of range).}
    \label{fig:perfect_heat_flow}
\end{figure}
~\cref{fig:perfect_heat_flow}a, b show the heat flow normalized to the corresponding free-space value $\dot{q}_{\mathrm{free}}$ as a function of cavity length for several values of $\Omega$ and for both dipole orientations. For a dipole oriented parallel to the cavity mirrors, the heat flow in an ideal cavity is suppressed, rather than enhanced, at short cavity lengths. On the contrary, for perpendicular dipole orientations, heat transfer remains enhanced in an ideal cavity. Nevertheless, the enhancement can be substantially larger in a cavity with gold as opposed to perfect mirrors. This result, as already mentioned in the main text, can be attributed to the contribution of evanescent modes, which are absent in the perfect mirror case. For example, for an oscillator frequency $\Omega/(2\pi) = 100$ GHz and at a cavity length $ L = 100$ nm, the heat flow for a dipole perpendicular to the mirrors is enhanced by a factor of approximately $3 \times 10^3$ with respect to the free-space heat flow for the perfect cavity, whereas the corresponding enhancement for the real cavity is approximately $10^5$. This comparison highlights the significant contribution of the evanescent component to the heat flow in the regime where the cavity resonance is tuned far above the oscillator frequency $\Omega$ (i.e. where $\nu_c \gg \Omega/(2\pi)$).

\bibliography{MyLibrary}